\documentclass[aps,preprint,showpacs]{revtex4}

\def\vb#1{\mbox{\boldmath$#1$}}
\def\pd#1#2{\frac{\partial #1}{\partial #2}}

\def\bdot{\,\vb{\cdot}\,}
\def\btimes{\,\vb{\times}\,}

\newcommand{\bc}{\begin{center}}
\newcommand{\ec}{\end{center}}
\newcommand{\bt}{\begin{tabbing}}
\newcommand{\et}{\end{tabbing}} 
\newcommand{\be}{\begin{eqnarray*}}
\newcommand{\ee}{\end{eqnarray*}}

\begin{document}

\title{Noether Methods for Fluids and Plasmas}
\author{A.J.~Brizard}
\affiliation{Department of Physics, Saint Michael's College \\
                One Winooski Park, Colchester, VT 05439, USA}

\date{October 13, 2004}

\begin{abstract}

Two applications of the Noether method for fluids and plasmas are presented based on the Euler-Lagrange and Euler-Poincar\'{e} variational principles, which depend on whether the dynamical fields are to be varied independently or not, respectively. The relativistic cold laser-plasma equations, describing the interaction between an intense laser field with a cold relativistic electron plasma, provide a useful set of equations amenable to both variational formulations. The derivation of conservation laws by Noether method proceeds from the Noether equation, whose form depends on the variational formulation used. As expected, the expressions for the energy-momentum conservation laws are identical in both variational formulations. The connection between the two Lagrangian densities is shown to involve the mass conservation and Lin constraints associated with the cold relativistic electron fluid.

\end{abstract}


\maketitle

\pagebreak

\section{Introduction}

The Noether method \cite{Goldstein} is used in many areas of physics to relate symmetry properties of Lagrangians to important conservation laws for dissipationless dynamical equations such as energy, momentum, and wave action. Most conventional applications of the Noether method in classical and quantum field theories \cite{Ramond,Morse_Feshbach,Yougrau_Mandelstam} focus on a variational formulation that considers independent variations of all dynamical fields. The purpose of this paper is to show how the Noether method can be applied to derive the energy-momentum conservation law for a set of dissipationless plasma-fluid equations from two different variational formulations, which differ by their treatment of field variations. 

We begin with an expression for a typical action functional in field theory \cite{Ramond}:
\begin{equation} 
{\cal A}[\psi_{a}] \;=\; \int_{{\cal D}} d^{4}x\; {\cal L}(\psi_{a},\;\partial_{\mu}\psi_{a}),
\label{eq:field_action}
\end{equation}
where the multi-component field $\psi_{a}({\bf x},t)$ represents the state of the system at position ${\bf x}$ and time 
$t$ (the index $a$ is used to represent a field component) and we assume that the multi-component field is constrained to the space-time domain ${\cal D}$ in the course of its evolution. In the present Section, a covariant notation is used in terms of the Minkowski metric tensor $\eta^{\mu\nu} = {\rm diag}(1,-1,-1,-1)$, the infinitesimal space-time volume is $d^{4}x = c dt\,d^{3}x$, and $\partial_{\mu} = (c^{-1}\partial_{t},\; \nabla)$ is the covariant space-time gradient. The field equation for the dissipationless time evolution of the multi-component field $\vb{\psi}({\bf x},t)$ is derived from a variational principle based on variations of the action functional (\ref{eq:field_action}):
\begin{eqnarray} 
\delta{\cal A}[\psi_{a}] & = & {\cal A}[\psi_{a} + \delta\psi_{a}] \;-\; {\cal A}[\psi_{a}] \;=\; \int_{{\cal D}}\;
\delta{\cal L}(\psi_{a},\;\partial_{\mu}\psi_{a})\; d^{4}x \nonumber \\
 & = & \sum_{a}\;\int_{{\cal D}} \left[\; \pd{{\cal L}}{\psi_{a}}\;\delta\psi_{a} \;+\; 
\pd{{\cal L}}{(\partial_{\mu}\psi_{a})}\;\pd{\delta\psi_{a}}{x^{\mu}} \;\right]\;d^{4}x,
\label{eq:deltaA_def}
\end{eqnarray}
where the variations $\delta\psi_{a}$ are assumed to vanish on the boundary $\partial{\cal D}$ of the space-time domain ${\cal D}$. 

The stationarity of the action leads to the variational principle
\begin{equation}
\delta{\cal A}[\psi_{a}] \;=\; \int_{{\cal D}}\; \delta{\cal L}(\psi_{a},\;\partial_{\mu}\psi_{a})\;d^{4}x \;=\; 0.
\label{eq:action_vp}
\end{equation}
The outcome of the variational principle (\ref{eq:action_vp}) is a set of partial differential equations involving the Lagrangian density ${\cal L}$ and its partial derivatives while the specific form of these equations depends on whether the variations $\delta\psi_{a}$ are independent or not. These two cases lead to the Euler-Lagrange equations 
\cite{Morse_Feshbach} and the Euler-Poincar\'{e} equations \cite{HMR}, respectively. As a result of the variational principle (\ref{eq:action_vp}), the variation 
$\delta{\cal L}$ of the Lagrangian density becomes the Noether equation
\begin{equation}
\delta{\cal L} \;\equiv\; \partial_{\mu}\,\Lambda^{\mu},
\label{eq:deltaL_general}
\end{equation}
where the four-density $\Lambda^{\mu}$ is expressed in terms of partial derivatives of the Lagrangian density ${\cal L}$. The Noether method associates symmetries of the Lagrangian density ${\cal L}$ with local conservation laws as will be shown below.

The remainder of the paper is organized as follows. In Sec.~II, the standard application of the Noether method based on an Euler-Lagrange variational principle is reviewed. A general expression for the Euler-Lagrange form of the energy-momentum conservation law is derived by Noether method based on symmetries of the Euler-Lagrange Lagrangian density with respect to space-time translations. Some of the invariance properties of the energy-momentum conservation law are also investigated within the context of the symmetry of the energy-momentum tensor. 

In Sec.~III, the relativistic cold laser-plasma equations \cite{lpi} describing the interaction between an intense laser field with a cold relativistic electron plasma in the presence of a neutralizing fixed-ion background. This example provides a useful set of equations amenable to both Euler-Lagrange and Euler-Poincar\'{e} variational formulations and each application of the Noether method leads to the same energy-momentum conservation law. In the Euler-Lagrange variational formulation, the dynamical fields $(\varphi, {\bf A}; n,\psi,\alpha,\beta)$ include the electromagnetic potentials $(\varphi, {\bf A})$, the electron density $n$, and the Clebsch potentials $(\psi, \alpha, \beta)$ used to represent the relativistic electron canonical momentum ${\bf P} \equiv \nabla\psi + \alpha\,\nabla\beta$ (in the present work, the curl-free case ${\bf P} = \nabla\psi$ discussed in Ref.~\cite{lpi} is generalized by considering 
$\nabla\btimes{\bf P} = \nabla\alpha\btimes\nabla\beta \neq 0$). In the Euler-Poincar\'{e} variational formulation, on the other hand, the Clebsch potentials $(\psi, \alpha, \beta)$ are replaced by the electron fluid velocity ${\bf v}$. We also show that these two Lagrangian densities are in fact connected by a simple formula. Lastly, we summarize our work in Sec.~V.

\section{CONSERVATION LAWS OF EULER-LAGRANGE EQUATIONS}

Standard Euler-Lagrange (EL) field equations are derived from the variational principle (\ref{eq:action_vp}) when the variations $\delta\psi_{a}$ are all independent. The variation $\delta{\cal L}$ of the Lagrangian density in Eq.~(\ref{eq:deltaA_def}) is, thus, explicitly written as
\begin{equation}
\delta{\cal L} \;=\; \sum_{a}\; \delta\psi_{a} \left\{\; \pd{{\cal L}}{\psi_{a}} \;-\; \pd{}{x^{\mu}} \left[ \pd{{\cal L}}{(\partial_{\mu}\psi_{a})} \right] \;\right\} \;+\; \pd{\Lambda^{\mu}_{{\rm EL}}}{x^{\mu}},
\label{eq:deltaL_EL}
\end{equation}
where the exact space-time divergence is obtained by rearranging terms with
\begin{equation} 
\Lambda^{\mu}_{{\rm EL}} \;=\; \sum_{a}\; \delta\psi_{a}\; \left[\; \pd{{\cal L}}{(\partial_{\mu}\psi_{a})} \;\right].
\label{eq:Lambda_def}
\end{equation}
The stationarity of the action functional associated with the variational principle (\ref{eq:action_vp}) yields
\begin{equation} 
0 \;=\; \sum_{a}\; \int_{\cal D} d^{4}x\; \delta\psi_{a} \left\{\; \pd{{\cal L}}{\psi_{a}} \;-\; \pd{}{x^{\mu}} \left[ \pd{{\cal L}}{(\partial_{\mu}\psi_{a})} \right] \;\right\},
\label{eq:station_action}
\end{equation}
where the exact space-time divergence $\partial_{\mu}\Lambda^{\mu}$ drops out under the assumption that the variations 
$\delta\psi_{a}$ vanish on $\partial{\cal D}$. 

Following the standard rules of Calculus of Variations 
\cite{Morse_Feshbach}, the condition that Eq.~(\ref{eq:station_action}) holds for any variation $\delta\psi_{a}$ yields the Euler-Lagrange equation for the field-component $\psi_{a}$:
\begin{equation} 
\pd{}{x^{\mu}} \left[\; \pd{{\cal L}}{(\partial_{\mu}\psi_{a})} \;\right] \;=\; \pd{}{t} \left[
\pd{{\cal L}}{(\partial_{t}\psi_{a})} \;\right] \;+\; \nabla\bdot\left[\; 
\pd{{\cal L}}{(\nabla\psi_{a})} \;\right] \;=\; \pd{{\cal L}}{\psi_{a}}.
\label{eq:EL_field}
\end{equation} 
Since the Euler-Lagrange equations (\ref{eq:EL_field}) hold for arbitrary variations $\delta\psi_{a}$, the variational equation (\ref{eq:deltaL_EL}) becomes 
\begin{equation} 
\delta{\cal L} \;\equiv\; \pd{\Lambda^{\mu}_{{\rm EL}}}{x^{\mu}} \;=\; \pd{}{x^{\mu}} \left[\; \sum_{a}\;\delta\psi_{a}\;
\pd{{\cal L}}{(\partial_{\mu}\psi_{a})} \;\right],
\label{eq:noether_eq}
\end{equation}
which is the Euler-Lagrange form of the Noether equation.

The standard application of the Noether method involves the derivation of the energy-momentum conservation law. For this purpose, we consider arbitrary space-time translations ($x^{\nu} \rightarrow x^{\nu} + \delta x^{\nu}$) generated by the space-time displacement $\delta x^{\nu}$. Under this space-time transformation, the variations of the field-components and the Lagrangian density are expressed as 
\begin{equation}
\left. \begin{array}{rcl}
\delta\psi_{a}  & \equiv & -\;\delta x^{\nu}\;\partial_{\nu}\psi_{a} \\
\\
\delta{\cal L} & \equiv & -\;\partial_{\nu}(\delta x^{\nu}\;{\cal L}) \;+\; \delta x^{\nu}\;
\partial_{\nu}^{\prime}{\cal L}
\end{array} \right\},
\label{eq:field_em}
\end{equation}
where $\partial_{\nu}^{\prime}{\cal L} \equiv (\partial_{\nu}{\cal L})|_{\psi_{a}}$ denotes the space-time derivative of ${\cal L}$ at constant field $\psi_{a}$. The energy-momentum conservation law is written in terms of the energy-momentum tensor
\begin{equation} 
T^{\mu}_{\;\;\nu} \;\equiv\; {\cal L}\;\eta^{\mu}_{\;\;\nu} \;-\; \sum_{a}\;\pd{{\cal L}}{(\partial_{\mu}\psi_{a})}\;\partial_{\nu}\psi_{a}.
\label{eq:em_tensor}
\end{equation}
as
\begin{equation}
\partial_{\mu}T^{\mu}_{\;\;\nu} \;=\; \partial^{\prime}_{\nu}{\cal L}.
\label{eq:em_conservation}
\end{equation}
Hence, when the Lagrangian density ${\cal L}$ is independent of the space-time coordinate $x^{\lambda}$ (i.e., 
$\partial_{\lambda}^{\prime}{\cal L} = 0$), Noether's theorem yields the energy-momentum conservation law 
$\partial_{\mu}T^{\mu}_{\;\;\lambda} = 0$.

When the background medium is time independent, the energy conservation law is written as $\partial{\cal E}/\partial t + \nabla\bdot{\bf S} = 0$, where the energy density ${\cal E}$ and the energy-density flux ${\bf S}$ are defined as
\begin{equation} 
{\cal E} \;=\; \sum_{a}\; \left[ \pd{{\cal L}}{(\partial_{t}\psi_{a})}\;\pd{\psi_{a}}{t} \right] \;-\; {\cal L} \;\;\;{\rm and}\;\;\;
{\bf S} \;=\; \sum_{a}\; \left[ \pd{{\cal L}}{(\nabla\psi_{a})}\;\pd{\psi_{a}}{t} \right].
\label{eq:energy_noether}
\end{equation}
When the background medium is homogeneous, on the other hand, the momentum conservation law is written as $\partial
\vb{\Pi}/\partial t + \nabla\bdot{\sf T} = 0$, where the momentum density $\vb{\Pi}$ and the momentum-stress tensor 
${\sf T}$ are defined as
\begin{equation} 
\vb{\Pi} \;=\; -\;\sum_{a}\; \left[ \pd{{\cal L}}{(\partial_{t}\psi_{a})}\;\nabla \psi_{a} \right] \;\;\;{\rm and}\;\;\; {\sf T} \;=\; {\cal L}\;{\bf I} \;-\; \sum_{a}\; \left[ \pd{{\cal L}}{(\nabla\psi_{a})}\;\nabla\psi_{a} \right].
\label{eq:momentum_noether}
\end{equation}
Note that the energy conservation law is invariant under the transformation
\begin{equation}
({\cal E},\;{\bf S}) \;\rightarrow\; \left({\cal E} \;+\; \nabla\bdot{\bf C},\; {\bf S} \;-\; \pd{{\bf C}}{t}\right),
\label{eq:energy_gauge}
\end{equation}
where the vector ${\bf C}$ is an arbitrary function of space and time. The momentum conservation law, on the other hand, is invariant under the transformation
\begin{equation}
(\vb{\Pi},\;{\sf T}) \;\rightarrow\; \left(\vb{\Pi} \;+\; \nabla\bdot{\sf D},\; {\sf T} \;-\; \pd{{\sf D}}{t}\right),
\label{eq:momentum_gauge}
\end{equation}
where the tensor ${\sf D}$ is an arbitrary function of space and time. The invariance properties of the energy-momentum conservation law ($T^{\mu}_{\;\;\nu} \rightarrow T^{\mu}_{\;\;\nu} + \partial_{\sigma}K^{\sigma\mu}_{\nu}$, where 
$K^{\mu\sigma}_{\nu} = -\,K^{\sigma\mu}_{\nu}$ so that $\partial^{2}_{\mu\sigma}K^{\sigma\mu}_{\nu} = 0$) can be used, for example, to ensure that the momentum-stress tensor ${\sf T}$ is symmetric.

\section{RELATIVISTIC COLD LASER-PLASMA INTERACTIONS}

The basic set of nonlinear equations describing the interaction of an intense laser pulse with a cold relativistic electron plasma \cite{lpi} is given as
\begin{eqnarray}
\pd{n}{t} \;+\; \nabla\bdot n{\bf v} & = & 0, \label{eq:lp_continuity} \\
\pd{{\bf p}}{t} \;+\; {\bf v}\bdot\nabla{\bf p} & = & -\;e\; \left( {\bf E} \;+\; \frac{{\bf v}}{c}\btimes{\bf B} 
\right), \label{lp_motion} \\
\nabla\bdot{\bf E} & = & 4\pi\,e\; (N - n), \label{eq:lp_Poisson} \\
\nabla\btimes{\bf B} \;-\; \frac{1}{c}\;\pd{{\bf E}}{t} & = & -\;4\pi\,e\;n\;\frac{{\bf v}}{c}, \label{eq:lp_Ampere}
\end{eqnarray}
where ${\bf p} = m\gamma\,{\bf v}$ is the relativistic electron kinetic momentum, with the relativistic factor 
\begin{equation}
\gamma \;=\; \left( 1 - \frac{|{\bf v}|^{2}}{c^{2}} \right)^{-\frac{1}{2}} \;=\; \left( 1 + 
\frac{|{\bf p}|^{2}}{m^{2}c^{2}} \right)^{\frac{1}{2}}, 
\label{eq:gamma_def}
\end{equation}
while ${\bf E} = -\,\nabla\varphi - c^{-1}\partial_{t}{\bf A}$ and ${\bf B} = \nabla\btimes{\bf A}$ are the electric and magnetic fields expressed in terms of the electromagnetic potentials $(\varphi,{\bf A})$. In Eq.~(\ref{eq:lp_Poisson}), the neutralizing fixed-ion density $N$ is assumed to be nonuniform $(\nabla N \neq 0)$ but is taken to be time-independent (otherwise, the ion continuity equation would require ion flow). We note that, while the vector potential ${\bf A}$ includes both the laser field (${\bf A}_{0}$) and the field $({\bf A}_{p}$) induced by the plasma current response, the scalar potential $\varphi$ represents only the field $(\varphi_{p})$ induced by the plasma charge-separation response. Furthermore, we note that the assumption of a cold relativistic electron plasma is justified by the large electron {\it quiver} velocities ${\bf v}_{q} = -\,e {\bf A}_{0}/(m_{e}c)$ generated by the intense laser field, which dominate over the random thermal motion of electrons. Although thermal effects can easily be incorporated within a Lagrangian formulation, they represent unecessary complications for our present purposes.

We now show that this set of equations can be represented either in terms of an Euler-Lagrange variational formulation or an Euler-Poincar\'{e} variational formulation. The Euler-Lagrange formulation presented below focuses its attention on the relativistic electron canonical momentum 
${\bf P} = {\bf p} - e{\bf A}/c$ instead of the relativistic electron kinetic momentum ${\bf p}$, which satisfies the relativistic canonical momentum equation
\begin{equation} 
\pd{{\bf P}}{t} \;=\; \nabla\left( e\,\varphi \;-\; \gamma\,mc^{2}\right) \;+\; {\bf v}\btimes\nabla\btimes{\bf P},
\label{eq:momentum_dyn}
\end{equation}
obtained from Eq.~(\ref{lp_motion}). From Eq.~(\ref{eq:momentum_dyn}), we note that the dynamical evolution of the momentum vorticity $\vb{\Omega} \equiv \nabla\btimes{\bf P}$ is expressed as $\partial\vb{\Omega}/\partial t = \nabla\btimes({\bf v}\btimes\vb{\Omega})$, which implies that the evolution of 
$\vb{\Omega}$ requires that its initial value be nonzero (i.e., if the electron canonical momentum is initially curl-free, it remains so during its subsequent evolution). The Euler-Poincar\'{e} formulation, on the other hand, considers variations in electron density 
$n$ and fluid velocity ${\bf v}$ to be constrained by the conservation of mass, i.e., the variations $\delta n$ and $\delta{\bf v}$ must satisfy the perturbed continuity equation 
\begin{equation}
\left. \left. \pd{\delta n}{t} \;+\; \nabla\bdot \right( \delta n\,{\bf v} + n\,\delta{\bf v}\right) \;=\; 0.
\label{eq:lp_pertcont}
\end{equation}
Both variational formulations yield identical energy-momentum conservation laws. 

The energy conservation law is, first, written in {\it primitive} form as
\begin{eqnarray}
0 & = & \pd{}{t} \left[\; \frac{1}{4\pi}\;({\bf E} \;+\; \nabla\varphi)\bdot{\bf E} \;-\; \frac{1}{8\pi}\;\left( 
|{\bf E}|^{2} \;-\; |{\bf B}|^{2} \right) \;-\; e\,(n - N)\;\varphi \;+\; n \;\gamma\;mc^{2} \;\right] \nonumber \\
 & + & \nabla\bdot\left[\; n{\bf v}\;\left( -\,e\,\varphi \;+\; \gamma\;mc^{2}\right) \;-\; \frac{1}{4\pi}\;{\bf E}\,
\pd{\varphi}{t} \;+\; \frac{c}{4\pi}\;({\bf E} \;+\; \nabla\varphi)\btimes{\bf B}) \;\right].
\label{eq:energy_primitive}
\end{eqnarray}
Next, by removing the gauge vector ${\bf C} = {\bf E}\,\varphi/4\pi$ associated with the energy-gauge condition 
(\ref{eq:energy_gauge}), we obtain
\begin{eqnarray*}
0 & = & \pd{}{t} \left[\; n \;\gamma\;mc^{2} \;+\; \frac{1}{8\pi}\; \left( |{\bf E}|^{2} \;+\; |{\bf B}|^{2} \right) \;-\; \varphi \left( e\,(n - N) \;+\; 
\frac{\nabla\bdot{\bf E}}{4\pi} \right) \;\right] \\
 & + & \nabla\bdot\left[\; n{\bf v}\;\gamma\;mc^{2} \;+\; \frac{c}{4\pi}\;{\bf E}\btimes{\bf B} \;+\; \varphi \left( \frac{1}{4\pi}\;\pd{{\bf E}}{t}
\;-\; \frac{c}{4\pi}\;\nabla\btimes{\bf B} \;-\; e\,n\,{\bf v} \right) \;\right].
\end{eqnarray*}
Lastly, making use of the Maxwell equations (\ref{eq:lp_Poisson}) and (\ref{eq:lp_Ampere}) and the continuity equation 
(\ref{eq:lp_continuity}), we obtain the energy conservation law
\begin{equation}
0 \;=\; \pd{{\cal E}}{t} \;+\; \nabla\bdot{\bf S},
\label{eq:lp_energycon}
\end{equation}
where the energy density ${\cal E}$ and energy-density flux ${\bf S}$ are
\begin{eqnarray}
{\cal E} & = & n \;\gamma\;mc^{2} \;+\; \frac{1}{8\pi}\; \left( |{\bf E}|^{2} \;+\; |{\bf B}|^{2} \right), \label{eq:lp_energydens} \\
{\bf S} & = & n{\bf v}\;\gamma\;mc^{2} \;+\; \frac{c}{4\pi}\;{\bf E}\btimes{\bf B}. \label{eq:lp_energydensflux}
\end{eqnarray}

The momentum conservation is written in {\it primitive} form as
\begin{eqnarray}
-\; e\varphi\;\nabla N & = & \pd{}{t} \left[\; n\;\left( m\,\gamma{\bf v} \;-\; \frac{e}{c}\;{\bf A} \right) \;+\; 
\frac{{\bf E}\btimes{\bf B}}{4\pi c} \;-\; {\bf A}\;\left( \frac{\nabla\bdot{\bf E}}{4\pi c} \right) \;+\;
\nabla\bdot\left( \frac{{\bf E}\,{\bf A}}{4\pi c} \right) \;\right] \nonumber \\
 & + & \nabla\bdot \left[\; n{\bf v}\;\left( m\,\gamma{\bf v} \;-\; \frac{e}{c}\;{\bf A} \right) \;-\; \frac{1}{4\pi} \left( {\bf E}\,{\bf E} \;+\; {\bf B}\,{\bf B}\right) \;+\; \frac{1}{8\pi}\;\left( |{\bf E}|^{2} \;+\; |{\bf B}|^{2} \right)\;{\bf I} \;\right. \nonumber \\
 &   &\left.-\; \pd{}{t} \left( \frac{{\bf E}\,{\bf A}}{4\pi c} \right) \;-\; e\,N\;\varphi\;{\bf I} \;+\; 
\left( \frac{1}{c}\;\pd{{\bf E}}{t} \;-\; \nabla\btimes{\bf B} \right)\;\frac{{\bf A}}{4\pi} \;\right],
\label{eq:momentum_primitive}
\end{eqnarray}
We now remove the gauge tensor ${\sf D} = {\bf E}\,{\bf A}/4\pi c$ associated with the momentum-gauge condition 
(\ref{eq:momentum_gauge}) to obtain
\begin{eqnarray*}
-\; e\varphi\;\nabla N & = & \pd{}{t} \left[\; n\;\left( m\,\gamma{\bf v} \;-\; \frac{e}{c}\;{\bf A} \right) \;+\; 
\frac{{\bf E}\btimes{\bf B}}{4\pi c} \;-\; {\bf A}\;\left( \frac{\nabla\bdot{\bf E}}{4\pi c} \right) \;\right] \\
 & + & \nabla\bdot \left[\; n{\bf v}\;\left( m\,\gamma{\bf v} \;-\; \frac{e}{c}\;{\bf A} \right) \;-\; \frac{1}{4\pi} \left( {\bf E}\,{\bf E} \;+\; 
{\bf B}\,{\bf B}\right) \;+\; \frac{1}{8\pi}\;\left( |{\bf E}|^{2} \;+\; |{\bf B}|^{2} \right)\;{\bf I} \;\right. \nonumber \\
 &   &\left.-\; e\,N\;\varphi\;{\bf I} \;+\; \left( \frac{1}{c}\;
\pd{{\bf E}}{t} \;-\; \nabla\btimes{\bf B} \right)\;\frac{{\bf A}}{4\pi} \;\right],
\end{eqnarray*}
Next, making use of the Maxwell equations (\ref{eq:lp_Poisson}) and (\ref{eq:lp_Ampere}), we obtain
\begin{eqnarray*}
-\; e\varphi\;\nabla N & = & \pd{}{t} \left[\; n\;m\,\gamma{\bf v} \;+\; \frac{{\bf E}\btimes{\bf B}}{4\pi c} \;-\; \frac{e\,N}{c}\;{\bf A} \;\right] \\
 & + & \nabla\bdot \left[\; nm\,\gamma\;{\bf v}{\bf v} \;-\; \frac{1}{4\pi} \left( {\bf E}\,{\bf E} \;+\; {\bf B}\,{\bf B}\right) \;+\; 
\frac{1}{8\pi}\;\left( |{\bf E}|^{2} \;+\; |{\bf B}|^{2} \right)\;{\bf I} \;-\; e\,N\;\varphi\;{\bf I} \;\right],
\end{eqnarray*}
which finally becomes
\begin{equation}
-\; eN\;{\bf E} \;=\; \pd{\vb{\Pi}}{t} \;+\; \nabla\bdot{\sf T},
\label{eq:lp_momentumcon}
\end{equation}
where the momentum density $\vb{\Pi}$ and the symmetric momentum-stress tensor ${\sf T}$ are
\begin{eqnarray}
\vb{\Pi} & = & n\;m\,\gamma{\bf v} \;+\; \frac{{\bf E}\btimes{\bf B}}{4\pi c}, \label{eq:lp_momentumdens} \\
{\sf T} & = & nm\,\gamma\;{\bf v}{\bf v} \;-\; \frac{1}{4\pi} \left( {\bf E}\,{\bf E} \;+\; {\bf B}\,{\bf B}\right) \;+\; 
\frac{1}{8\pi}\;\left( |{\bf E}|^{2} \;+\; |{\bf B}|^{2} \right)\;{\bf I}. \label{eq:lp_momentumstress_tensor}
\end{eqnarray}
Hence, Eq.~(\ref{eq:lp_momentumcon}) implies that the total momentum is conserved in the direction where the electric field vanishes. Note that, while the momentum-stress tensor in the primitive form of the momentum conservation law 
(\ref{eq:momentum_primitive}) is not symmetric, the final form of the momentum-stress tensor (\ref{eq:lp_momentumstress_tensor}) is symmetric.

\subsection{Euler-Lagrange Formulation}

The Euler-Lagrange formulation of the relativistic cold laser-plasma equations (\ref{eq:lp_continuity})-(\ref{eq:lp_Ampere}) is based on the use of Clebsch (or Euler) potentials \cite{SW}. According to the Clebsch representation associated with the general vorticity condition $(\vb{\Omega} = \nabla\btimes{\bf P} \neq 0)$, the relativistic electron canonical momentum is expressed as 
\begin{equation}
{\bf P} \;=\; m\gamma\,{\bf v} \;-\; \frac{e}{c}\;{\bf A} \;\equiv\; \nabla\psi \;+\; \alpha\;\nabla\beta,
\label{eq:momentum_lp}
\end{equation}
where the Clebsch potentials $\psi$, $\alpha$, $\beta$ are functions of space and time. As will be shown below, the potential $\psi$ may be viewed as a Lagrange multiplier associated with the mass conservation law (e.g., the continuity equation), while the potentials $\alpha$ and $\beta$ ensure that $\nabla\btimes{\bf P} = \nabla\alpha\btimes\nabla\beta \neq 0$. Furthermore, the Lin constraints \cite{SW} on the Clebsch potentials $\alpha$ and $\beta$ requires that the potentials $\alpha$ and $\beta$ be carried by the fluid flow in the sense that 
\begin{equation}
\pd{\alpha}{t} \;+\; {\bf v}\bdot\nabla\alpha \;=\; 0 \;=\; \pd{\beta}{t} \;+\; {\bf v}\bdot\nabla\beta,
\label{eq:Lin}
\end{equation}
i.e., the potentials $\alpha$ and $\beta$ are known as Lagrangian-{\it marker} coordinates \cite{Brown,Kats}. Note that under a gauge transformation $(\varphi,\, {\bf A}) \rightarrow (\varphi - c^{-1}\partial_{t}\chi,\, {\bf A} + 
\nabla\chi)$, the Clebsch potential $\psi$ obeys the gauge transformation $\psi \rightarrow \psi - (e/c)\,\chi$ while the Lagrangian-marker coordinates $\alpha$ and $\beta$ are invariant. Lastly, a more symmetric expression for the Clebsch representation of the relativistic canonical momentum can also be obtained by writing Eq.~(\ref{eq:momentum_lp}) as 
\begin{equation} 
{\bf P} \;=\; \nabla\psi^{\prime} \;+\; \frac{1}{2}\;(\alpha\,\nabla\beta \;-\; \beta\,\nabla\alpha),
\label{eq:mom_sym}
\end{equation}
which involves a redefinition of the Clebsch potential $\psi^{\prime} = \psi + \frac{1}{2}\;\alpha\beta$.

The Lagrangian density in the Euler-Lagrange formulation of the relativistic cold laser-plasma equations 
(\ref{eq:lp_continuity}), (\ref{eq:lp_Poisson})-(\ref{eq:lp_Ampere}), and (\ref{eq:momentum_dyn}) is
\begin{equation}
{\cal L}_{EL} \;=\; \frac{1}{8\pi}\;\left( |{\bf E}|^{2} \;-\; |{\bf B}|^{2} \right) \;+\; e\,(n - N)\;\varphi \;-\; n \left( \pd{\psi}{t} \;+\;
\alpha\;\pd{\beta}{t} \;+\; \gamma\;mc^{2} \right),
\label{eq:lp_EL}
\end{equation}
where the relativistic factor is
\begin{equation} 
\gamma \;=\; \sqrt{1 \;+\; \frac{1}{m^{2}c^{2}}\;\left| {\bf P} \;+\; \frac{e}{c}\;{\bf A} \right|^{2}},
\label{eq:lp_gamma}
\end{equation}
where the relativistic canonical momentum is expressed in terms of Clebsch potentials as given in 
Eq.~(\ref{eq:momentum_lp}). Here, the variational fields $(\varphi,{\bf A}; n,\psi,\alpha,\beta)$ are all to be varied independently in the variational principle 
\begin{equation}
\delta\int {\cal L}_{EL}\,d^{4}x = 0.
\label{eq:EL_vp}
\end{equation} 
The variation of the Lagrangian density (\ref{eq:lp_EL}) can be expressed as
\begin{eqnarray}
\delta{\cal L}_{EL} & = & \delta n \left( e\,\varphi \;-\; \pd{\psi}{t} \;-\; \alpha\;\pd{\beta}{t} \;-\; \gamma\;mc^{2} \right) \;+\; \left( 
\delta\psi \;+\; \alpha\;\delta\beta \right) \left( \pd{n}{t} \;+\; \nabla\bdot n{\bf v} \right) \nonumber \\
 & - & n\; \left[\; \delta\alpha \left( \pd{\beta}{t} \;+\; {\bf v}\bdot\nabla\beta \right) \;-\; \delta\beta \left( \pd{\alpha}{t} \;+\;
{\bf v}\bdot\nabla\alpha \right) \;\right] \nonumber \\
 & + & \left. \left. \frac{\delta\varphi}{4\pi} \right[\; 4\pi\,e\,(n - N) \;+\; \nabla\bdot{\bf E} \;\right] \;+\; \frac{\delta{\bf A}}{4\pi}\bdot 
\left( \frac{1}{c}\;\pd{{\bf E}}{t} \;-\; \nabla\btimes{\bf B} \;-\; 4\pi\,en\,\frac{{\bf v}}{c} \right) \nonumber \\
 & - & \pd{}{t} \left[\; n\;(\delta\psi \;+\; \alpha\;\delta\beta) \;+\; \frac{1}{4\pi c}\;\delta{\bf A}\bdot{\bf E} \;\right] \nonumber \\
 & - & \nabla\bdot \left[\; n{\bf v}\;(\delta\psi \;+\; \alpha\,\delta\beta) \;+\; \frac{1}{4\pi}\;(\delta\varphi\,{\bf E} \;+\; \delta{\bf A}\btimes
{\bf B}) \;\right],
\label{eq:lpEL_var}
\end{eqnarray}
where we have rearranged terms in order to isolate variations in the dynamical variables $(\varphi,{\bf A}; n,\psi,\alpha,\beta)$ and, thus, extracted the Noether space-time divergence. 

As a result of the variational principle (\ref{eq:EL_vp}), the relativistic cold plasma-laser equations (\ref{eq:lp_continuity}), 
(\ref{eq:lp_Poisson})-(\ref{eq:lp_Ampere}), and (\ref{eq:Lin}) are easily recovered. The last variational equation, which corresponds to the Euler-Lagrange equation $\partial{\cal L}_{EL}/\partial n = 0$, is the energy equation
\begin{equation}
\pd{\psi}{t} \;+\; \alpha\;\pd{\beta}{t} \;=\; e\,\varphi \;-\; \gamma\;mc^{2}.
\label{eq:EL_energy}
\end{equation}
We recover the relativistic electron canonical momentum equation (\ref{eq:momentum_dyn}) by first taking the gradient of Eq.~(\ref{eq:EL_energy}) and rearranging terms to obtain
\[ \partial_{t}(\nabla\psi \;+\; \alpha\;\nabla\beta) \;=\; \nabla\left(e\varphi \;-\; \gamma\,mc^{2}\right) \;+\; {\bf v}\btimes\left(\nabla\alpha
\btimes\nabla\beta\right). \]
By substituting the Clebsch representation (\ref{eq:momentum_lp}), we easily obtain Eq.~(\ref{eq:momentum_dyn}). Since these variational equations hold for arbitrary variations in $(\varphi,{\bf A}; n,\psi,\alpha,\beta)$, the expression for the variation (\ref{eq:lpEL_var}) of the Lagrangian density can now be written as the Euler-Lagrange form of the Noether equation
\begin{eqnarray}
\delta{\cal L}_{EL} & = & -\; \pd{}{t} \left[\; n\;(\delta\psi \;+\; \alpha\;\delta\beta) \;+\; \frac{1}{4\pi c}\;\delta{\bf A}\bdot{\bf E} \;\right] \nonumber \\
 &  &\mbox{}-\; \nabla\bdot \left[\; n{\bf v}\;(\delta\psi \;+\; \alpha\,\delta\beta) \;+\; \frac{1}{4\pi}\;(\delta\varphi\,{\bf E} \;+\; \delta{\bf A}\btimes{\bf B}) \;\right],
\label{eq:lpEL_Noether}
\end{eqnarray}
which can now be used to derive the conservation laws of energy and momentum for the relativistic cold laser-plasma equations.

\subsubsection{Energy conservation law for the relativistic cold laser-plasma equations}

The conservation law of energy is derived from the Noether equation (\ref{eq:lpEL_Noether}) by considering time translations $t \rightarrow t + 
\delta t$, such that $\delta\psi = -\,\delta t\,\partial_{t}\psi$, $\delta\beta = -\,\delta t\,\partial_{t}\beta$, $\delta\varphi = -\,\delta t\,
\partial_{t}\varphi$, 
\[ \delta{\bf A} \;=\; -\,\delta t\,\partial_{t}{\bf A} \;=\; c\,\delta t \;({\bf E} \;+\; \nabla\varphi), \]
and $\delta{\cal L}_{EL} = -\,\delta t\,\partial_{t}{\cal L}_{EL}$ (where we use the fact that the fixed-ion density $N$ is time-independent). Substituting these expressions into Eq.~(\ref{eq:lpEL_Noether}), we first obtain
\begin{eqnarray*}
-\;\partial_{t}{\cal L}_{EL} & = & \pd{}{t} \left[\; n\;\left( \pd{\psi}{t} \;+\; \alpha\;\pd{\beta}{t} \right) \;-\; \frac{1}{4\pi}\;({\bf E} 
\;+\; \nabla\varphi)\bdot{\bf E} \;\right] \\
 & + & \nabla\bdot \left[\; n{\bf v}\;\left( \pd{\psi}{t} \;+\; \alpha\;\pd{\beta}{t} \right) \;+\; \frac{{\bf E}}{4\pi}\;\pd{\varphi}{t} 
\;-\; \frac{c}{4\pi}\;({\bf E} \;+\; \nabla\varphi)\btimes{\bf B}) \;\right],
\end{eqnarray*}
which, by making use of Eq.~(\ref{eq:EL_energy}), becomes the primitive form of the energy conservation law 
(\ref{eq:energy_primitive}) and, thus, yields the energy conservation law (\ref{eq:lp_energycon}).

\subsubsection{Momentum conservation law for the relativistic cold laser-plasma equations}

The conservation law of momentum is derived from the Noether equation (\ref{eq:lpEL_Noether}) by considering space translations ${\bf x} \rightarrow 
{\bf x} + \delta{\bf x}$, such that $\delta\psi = -\,\delta{\bf x}\bdot\nabla\psi$, $\delta\beta = -\,\delta{\bf x}\bdot\nabla\beta$, 
\[ \delta\varphi \;=\; -\,\delta{\bf x}\bdot\nabla\varphi \;=\; \delta{\bf x}\bdot\left( {\bf E} \;+\; \frac{1}{c}\;\pd{{\bf A}}{t} \right),\;\;\; 
\delta{\bf A} \;=\; -\,\delta{\bf x}\bdot\nabla{\bf A} \;=\; \delta{\bf x}\btimes{\bf B} \;-\; \nabla\left({\bf A}\bdot\delta{\bf x}\right), \]
and $\delta{\cal L}_{EL} = -\,\nabla\bdot(\delta{\bf x}\;{\cal L}_{EL}) + \delta{\bf x}\bdot\nabla^{\prime}{\cal L}_{EL}$ (where $\nabla^{\prime}
{\cal L}_{EL} = -\,e\varphi\;\nabla N$ for a nonuniform fixed-ion density). Substituting these expressions into 
Eq.~(\ref{eq:lpEL_Noether}), we first obtain
\begin{eqnarray*}
-\;\partial_{i}{\cal L}_{EL} \;-\; e\varphi\;\partial_{i}N & = & \pd{}{t} \left[\; n\;(\partial_{i}\psi \;+\; \alpha\;\partial_{i}\beta) \;+\; \left.
\left. \frac{1}{4\pi c}\;E^{j} \right( \epsilon_{ijk}\;B^{k} \;+\; \partial_{j}A_{i} \right) \;\right] \\
 & + & \partial_{j} \left[\; nv^{j}\;(\partial_{i}\psi \;+\; \alpha\,\partial_{i}\beta) \;-\; \frac{1}{4\pi} \left( E^{j}\,E_{i} \;+\; B^{j}\,B_{i}\right)
\right. \\
 & - &\left. \frac{E^{j}}{4\pi c}\;\pd{A_{i}}{t} \;+\; \frac{B_{k}}{4\pi} \left( B^{k}\;\delta^{j}_{\;i} \;-\; 
\epsilon^{jk\ell}\; \partial_{\ell}A_{i} \right) \;\right].
\end{eqnarray*}
Next, substituting the Clebsch representation (\ref{eq:momentum_lp}) for the relativistic canonical momentum, we obtain the primitive form (\ref{eq:momentum_primitive}) of the momentum conservation law and, thus, the momentum conservation law (\ref{eq:lp_momentumcon}) follows.

\subsection{Euler-Poincar\'{e} Formulation}

The Euler-Poincar\'{e} variational formulation of the relativistic cold laser-plasma equations 
(\ref{eq:lp_continuity})-(\ref{eq:lp_Ampere}) is expressed in terms of the Lagrangian density
\begin{equation}
{\cal L}_{EP} \;=\; \frac{1}{8\pi} \left( |{\bf E}|^{2} \;-\; |{\bf B}|^{2} \right) \;+\; e\,(n - N)\,\varphi \;-\; \frac{e}{c}\,n{\bf v}\bdot{\bf A} \;-\;
n\;\gamma^{-1}mc^{2},
\label{eq:lp_EP}
\end{equation}
where $\gamma^{-1} = (1 - |{\bf v}|^{2}/c^{2})^{\frac{1}{2}}$. The Eulerian variations of the electron density $n$ and fluid velocity ${\bf v}$, which are subject to the mass conservation constraint (\ref{eq:lp_pertcont}), are expressed in terms of the virtual fluid displacement $\delta\vb{\xi}$ as
\begin{equation}
\delta n \;=\; -\;\nabla\bdot( n\;\delta\vb{\xi}) \;\;\;{\rm and}\;\;\; \delta{\bf v} \;=\; \left( \pd{}{t} \;+\; {\bf v}\bdot\nabla \right)
\delta\vb{\xi} \;-\; \delta\vb{\xi}\bdot\nabla{\bf v}.
\label{eq:nv_var}
\end{equation}
Relativistically covariant expressions for the Eulerian variations for the proper density $N = \gamma^{-1}\,n$ and the fluid four-velocity $u^{\mu} = (\gamma c, \gamma{\bf v})$ are found in Ref.~\cite{BMW}; here, we prefer the Eulerian variations of $n$ and ${\bf v}$ for the sake of simplicity.

The variation of the Euler-Poincar\'{e} Lagrangian density can be expressed as
\begin{eqnarray}
\delta{\cal L}_{EP} & = &  \left. \left. \frac{\delta\varphi}{4\pi} \right[\; 4\pi\,e\,(n - N) \;+\; \nabla\bdot{\bf E} \;\right] \;+\; 
\frac{\delta{\bf A}}{4\pi}\bdot \left( \frac{1}{c}\;\pd{{\bf E}}{t} \;-\; \nabla\btimes{\bf B} \;-\; 4\pi\,en\,\frac{{\bf v}}{c} \right) \nonumber \\
 & + & n\,\delta\vb{\xi}\bdot \left[\; \nabla\left(e\,\varphi - \gamma\,mc^{2}\right) \;-\; \pd{{\bf P}}{t} \;+\; {\bf v}\btimes\nabla\btimes{\bf P} 
\;\right] \nonumber \\
 & + & \pd{}{t} \left( n\,{\bf P}\bdot\delta\vb{\xi} \;-\; \frac{1}{4\pi c}\;\delta{\bf A}\bdot{\bf E} \right) \;+\; \nabla\bdot \left[\; 
n\,\delta\vb{\xi}\;\left( \gamma^{-1}mc^{2} \;-\; e\,\varphi \;+\; \frac{e}{c}\,{\bf v}\bdot{\bf A} \right) \right. \nonumber \\
 & + &\left. n\,{\bf v}\;{\bf P}\bdot\delta\vb{\xi} \;-\; \left. \left. \frac{1}{4\pi}\;\right( \delta\varphi\,{\bf E} \;+\; \delta{\bf A}\btimes
{\bf B}\right) \;\right],
\label{eq:lpEP_var}
\end{eqnarray}
where ${\bf P} = m\,\gamma{\bf v} - (e/c)\,{\bf A}$ denotes the relativistic canonical momentum and we have rearranged terms in order to isolate variations in the dynamical variables $(\vb{\xi},\varphi,{\bf A})$ and, thus, extracted the Noether space-time divergence. 

As a result of the variational principle
\begin{equation}
\delta\;\int\;{\cal L}_{EP}\;d^{4}x \;=\; 0,
\label{eq:EP_vp}
\end{equation}
the relativistic cold laser-plasma equations (\ref{eq:lp_Poisson})-(\ref{eq:lp_Ampere}) and (\ref{eq:momentum_dyn}) are easily recovered (under arbitrary variations 
$\delta\varphi$, $\delta{\bf A}$, and $\delta\vb{\xi}$, respectively) with the continuity equation (\ref{eq:lp_continuity}) representing a constraint equation. Since these variational equations hold for arbitrary variations in $(\vb{\xi},\varphi,{\bf A})$, the expression for the variation 
(\ref{eq:lpEP_var}) of the Lagrangian density can now be written as the Euler-Poincar\'{e} form of the Noether equation
\begin{eqnarray}
\delta{\cal L}_{EP} & = & \pd{}{t} \left( n\,{\bf P}\bdot\delta\vb{\xi} \;-\; \frac{1}{4\pi c}\;\delta{\bf A}\bdot{\bf E} \right) \;+\; \nabla\bdot 
\left[\; n\,\delta\vb{\xi}\;\left( \gamma^{-1}mc^{2} \;-\; e\,\varphi \;+\; \frac{e}{c}\,{\bf v}\bdot{\bf A} \right) \right. \nonumber \\
 &  &\left.+\; n\,{\bf v}\;{\bf P}\bdot\delta\vb{\xi} \;-\; \left. \left. \frac{1}{4\pi}\;\right( \delta\varphi\,{\bf E} \;+\; \delta{\bf A}\btimes
{\bf B}\right) \;\right],
\label{eq:lpEP_Noether}
\end{eqnarray}
which can now be used to derive the conservation laws of energy and momentum for the relativistic cold plasma-laser equations. 

To derive the energy-momentum conservation laws within the Euler-Poincar\'{e} formulation, the virtual fluid displacement $\delta\vb{\xi}$ is expressed in terms of the space and time translations $\delta{\bf x}$ and $\delta t$ as
\begin{equation}
\delta\vb{\xi} \;=\; \delta{\bf x} \;-\; {\bf v}\;\delta t,
\label{eq:xi_em}
\end{equation}
where ${\bf v}$ is the fluid velocity. Note that the relativistic approach used here is based on constrained variations of the laboratory-frame electron fluid density and electron fluid velocity; an alternative approach is based on constrained variations of rest-frame fluid quantities.

\subsubsection{Energy conservation law for the relativistic cold laser-plasma equations}

The conservation law of energy is derived from the Noether equation (\ref{eq:lpEP_Noether}) by considering time translations $t \rightarrow t + 
\delta t$, such that $\delta\vb{\xi} = -\,{\bf v}\,\delta t$, $\delta\varphi = -\,\delta t\,\partial_{t}\varphi$, 
\[ \delta{\bf A} \;=\; -\,\delta t\,\partial_{t}{\bf A} \;=\; c\,\delta t \;({\bf E} \;+\; \nabla\varphi), \]
and $\delta{\cal L}_{EP} = -\,\delta t\,\partial_{t}{\cal L}_{EP}$ (where we use the fact that the fixed-ion density $N$ is time-independent). Substituting these expressions into Eq.~(\ref{eq:lpEP_Noether}), we first obtain
\begin{eqnarray*}
-\;\pd{{\cal L}_{EP}}{t} & = & -\; \pd{}{t} \left[\; n\,{\bf P}\bdot{\bf v} \;+\; \frac{1}{4\pi}\;({\bf E} \;+\; \nabla\varphi)\bdot{\bf E} \;\right] 
\;-\; \nabla\bdot\left[\; n\,{\bf v}\;{\bf P}\bdot{\bf v} \;-\; \frac{{\bf E}}{4\pi}\;\pd{\varphi}{t} \right. \\
 &   &\left.+\; n\,{\bf v}\;\left( \gamma^{-1}mc^{2} \;-\; e\,\varphi \;+\; \frac{e}{c}\,{\bf v}\bdot{\bf A} \right) \;+\; \frac{c}{4\pi}\;({\bf E} 
\;+\; \nabla\varphi)\btimes{\bf B} \;\right].
\end{eqnarray*}
Next, by substituting the expression
\begin{equation} 
{\bf P}\bdot{\bf v} \;=\; mc^{2} \left( \gamma \;-\; \gamma^{-1} \right) \;-\; \frac{e}{c}\,{\bf A}\bdot{\bf v},
\label{eq:Pdotv}
\end{equation}
we recover the primitive form (\ref{eq:energy_primitive}) of the energy conservation law, from which the energy conservation law (\ref{eq:lp_energycon}) follows. 

\subsubsection{Momentum conservation law for the relativistic cold laser-plasma equations}

The conservation law of momentum is derived from the Noether equation (\ref{eq:lpEP_Noether}) by considering space translations ${\bf x} \rightarrow 
{\bf x} + \delta{\bf x}$, such that $\delta\vb{\xi} = \delta{\bf x}$, 
\[ \delta\varphi \;=\; -\,\delta{\bf x}\bdot\nabla\varphi \;=\; \delta{\bf x}\bdot\left( {\bf E} \;+\; \frac{1}{c}\;\pd{{\bf A}}{t} \right),\;\;\; 
\delta{\bf A} \;=\; -\,\delta{\bf x}\bdot\nabla{\bf A} \;=\; \delta{\bf x}\btimes{\bf B} \;-\; \nabla\left({\bf A}\bdot\delta{\bf x}\right), \]
and $\delta{\cal L}_{EP} = -\,\nabla\bdot(\delta{\bf x}\;{\cal L}_{EP}) + \delta{\bf x}\bdot\nabla^{\prime}{\cal L}_{EP}$ (where $\nabla^{\prime}
{\cal L}_{EP} = -\,e\varphi\;\nabla N$ for a nonuniform fixed-ion density). Substituting these expressions into 
Eq.~(\ref{eq:lpEP_Noether}), we first obtain
\begin{eqnarray*}
 &  &-\;\partial_{i}{\cal L}_{EP} \;-\; e\varphi\;\partial_{i}N \;=\; \pd{}{t} \left[\; n\;P_{i} \;+\; \left. \left. \frac{1}{4\pi c}\;E^{j} \right( 
\epsilon_{ijk}\;B^{k} \;+\; \partial_{j}A_{i} \right) \;\right] \\
 &  &\mbox{}+\; \partial_{j} \left[\; nv^{j}\;P_{i} \;-\; \frac{1}{4\pi} \left( E^{j}\,E_{i} \;+\; B^{j}\,B_{i}\right) \;-\; \frac{E^{j}\,
\partial_{t}A_{i}}{4\pi c} \;+\; \frac{B_{k}}{4\pi} \left( B^{k}\;\delta^{j}_{\;i} \;-\; \epsilon^{jk\ell}\; \partial_{\ell}A_{i} \right) \;\right].
\end{eqnarray*}
Next, by substituting the expression ${\bf P} = m\gamma{\bf v} - (e/c)\,{\bf A}$ for the relativistic canonical momentum,
we recover the primitive form (\ref{eq:momentum_primitive}) of the momentum conservation law, from which the momentum conservation law (\ref{eq:lp_momentumcon}) follows.

\subsection{Connection between the Lagrangian Densities}

We have thus far shown that the relativistic cold laser-plasma equations (\ref{eq:lp_continuity})-(\ref{eq:lp_Ampere}) can be represented in terms of two different variational principles based on either the Euler-Lagrange Lagrangian density 
(\ref{eq:lp_EL}) or the Euler-Poincar\'{e} Lagrangian density (\ref{eq:lp_EP}). We now show the connection between these two Lagrangians. We begin by evaluating the difference:
\begin{equation} 
{\cal L}_{EL} \;-\; {\cal L}_{EP} \;=\; -\,n \left( \pd{\psi}{t} \;+\; \alpha\,\pd{\beta}{t} \right) \;-\; n\,mc^{2}
\left( \gamma \;-\; \gamma^{-1} \right) \;+\; \frac{e}{c}\,n{\bf v}\bdot{\bf A}.
\label{eq:diff_one}
\end{equation}
Next, by making use of Eq.~(\ref{eq:Pdotv}) with the Clebsch representation (\ref{eq:mom_sym}) for the canonical momentum 
${\bf P}$, the difference (\ref{eq:diff_one}) becomes 
\[ {\cal L}_{EL} \;-\; {\cal L}_{EP} \;=\; -\,n\; \left( \pd{\psi^{\prime}}{t} \;+\; {\bf v}\bdot\nabla\psi^{\prime} \right) \;-\; \frac{n}{2} \left[\; \alpha\; \left( \pd{\beta}{t} \;+\; {\bf v}\bdot\nabla\beta \right) \;-\;
\beta\; \left( \pd{\alpha}{t} \;+\; {\bf v}\bdot\nabla\alpha \right) \;\right]. \]
Here, the first term can be written as an exact space-time divergence plus a term involving the continuity equation and, thus, the difference ${\cal L}_{EL} - {\cal L}_{EP}$ is expressed as
\begin{eqnarray}
{\cal L}_{EL} \;-\; {\cal L}_{EP} & = & -\; \left[\; \pd{(n\psi^{\prime})}{t} \;+\; \nabla\bdot\left( 
n{\bf v}\,\psi^{\prime}\right) \;\right] \;+\; \psi^{\prime}\; \left( \pd{n}{t} \;+\; \nabla\bdot(n\,{\bf v}) \right) \nonumber \\
 &  &\mbox{}-\; \frac{n}{2} \left[\; \alpha\; \left( \pd{\beta}{t} \;+\; {\bf v}\bdot\nabla\beta \right) \;-\;
\beta\; \left( \pd{\alpha}{t} \;+\; {\bf v}\bdot\nabla\alpha \right) \;\right].
\label{eq:diff_two}
\end{eqnarray}
Lastly, since two Lagrangian densities that differ by an exact space-time divergence are considered equivalent (i.e., they generate identical dynamical equations), we may eliminate the exact space-time divergence from the right side of 
Eq.~(\ref{eq:diff_two}) to obtain the final expression
\begin{equation}
{\cal L}_{EL} \;=\; {\cal L}_{EP} \;+\; \psi^{\prime}\; \left( \pd{n}{t} \;+\; \nabla\bdot(n\,{\bf v}) \right) \;-\; 
\frac{n}{2} \left[\; \alpha\; \left( \pd{\beta}{t} \;+\; {\bf v}\bdot\nabla\beta \right) \;-\;
\beta\; \left( \pd{\alpha}{t} \;+\; {\bf v}\bdot\nabla\alpha \right) \;\right]
\label{eq:ELEP_connection}
\end{equation}
for the Euler-Lagrange Lagrangian density ${\cal L}_{EL}$ given in terms of the Euler-Poincar\'{e} Lagrangian density
${\cal L}_{EP}$, the mass-conservation constraint (\ref{eq:lp_continuity}) (with Lagrange multiplier $\psi^{\prime}$), and the Lin constraints (\ref{eq:Lin}) for the Lagrangian-marker coordinates $\alpha$ and $\beta$ (with Lagrange 
multipliers $n\beta/2$ and $-\,n\alpha/2$, respectively).

\section{Summary}

We have presented two variational formulations for the relativistic cold laser-plasma equations, which depend on whether all the dynamical fields are varied independently or not. We note that variational formulations in the physics literature are either presented in the Euler-Lagrange form or the Euler-Poincar\'{e} form (generally not both) and that the work presented here shows the explicit connection (\ref{eq:ELEP_connection}) between the two variational formulations for an important set of plasma-fluid equations. 

In deriving the energy-momentum conservation laws for the relativistic cold laser-plasma equations, we showed that, although the form of the Noether equation depends on the variational formulation used, i.e., either the Euler-Lagrange form (\ref{eq:lpEL_Noether}) or the Euler-Poincar\'{e} form (\ref{eq:lpEP_Noether}), the same conservation equations
(\ref{eq:lp_energycon}) and (\ref{eq:lp_momentumcon}) are obtained.

\vfill\eject

\end{document}